\documentclass[11pt,a4paper,english,nofootinbib]{revtex4}
\usepackage{lmodern}

\usepackage[T1]{fontenc}
\usepackage[latin9]{inputenc}
\usepackage{babel}
\usepackage{amsmath}
\usepackage{amssymb}
\usepackage{esint}
\usepackage[unicode=true,pdfusetitle,
 bookmarks=true,bookmarksnumbered=false,bookmarksopen=false,
 breaklinks=false,pdfborder={0 0 1},backref=false,colorlinks=false]
 {hyperref}

\makeatletter

\pdfpageheight\paperheight
\pdfpagewidth\paperwidth

\@ifundefined{textcolor}{}
{%
 \definecolor{BLACK}{gray}{0}
 \definecolor{WHITE}{gray}{1}
 \definecolor{RED}{rgb}{1,0,0}
 \definecolor{GREEN}{rgb}{0,1,0}
 \definecolor{BLUE}{rgb}{0,0,1}
 \definecolor{CYAN}{cmyk}{1,0,0,0}
 \definecolor{MAGENTA}{cmyk}{0,1,0,0}
 \definecolor{YELLOW}{cmyk}{0,0,1,0}
}

\usepackage{latexsym}\usepackage{bm}

\makeatother

\makeatother

\begin{document}

\title{Energy and Angular Momentum in Generic F(Riemann) Theories}

\author{Çetin \c{S}entürk}

\email{ctnsenturk@gmail.com}

\affiliation{no affiliation}

\author{Tahsin Ça\u{g}r\i{} \c{S}i\c{s}man}

\email{tahsin.c.sisman@gmail.com}

\affiliation{Department of Physics,\\
 Middle East Technical University, 06800 Ankara, Turkey}

\author{Bayram Tekin}

\email{btekin@metu.edu.tr}

\affiliation{Department of Physics,\\
 Middle East Technical University, 06800 Ankara, Turkey}

\date{\today}
\begin{abstract}
We construct the conserved charge of generic gravity theories built
on arbitrary contractions of the Riemann tensor (but not on its derivatives)
for asymptotically (anti)-de Sitter spacetimes. Our construction is
a generalization of the ADT charges of linear and quadratic gravity
theories in cosmological backgrounds. As an explicit example we find
the energy and angular momentum of the BTZ black hole in the 2+1 dimensional
Born-Infeld gravity. \tableofcontents{}
\[
\]

\end{abstract}
\maketitle

\section{Introduction}

For any geometric theory of gravity based on the Riemann tensor with
a Lagrangian density $\mathcal{L}\equiv\sqrt{-\left|g\right|}\, F\left(R_{\rho\sigma}^{\mu\nu}\right)$,
the conserved mass is given by the celebrated Arnowitt-Deser-Misner
(ADM) \cite{adm} formula for \emph{asymptotically flat} spaces: 
\begin{eqnarray}
M_{ADM}=\frac{1}{\kappa_{\text{Newton}}}\int_{S^{D-2}} & dS_{i} & \Big\{\partial_{j}h^{ij}-\partial^{i}h_{jj}\Big\},\label{eq:ADM_e}
\end{eqnarray}
where the perturbation is defined as $h_{\mu\nu}\equiv g_{\mu\nu}-\eta_{\mu\nu}$
and the integral is to be evaluated on a sphere at spatial infinity.
Note that the formula is written in terms of cartesian coordinates
even though it is a geometric invariant of the spatial part of the
spacetime manifold. For asymptotically flat spacetimes, angular momentum
(or momenta) has a similar expression. 
\begin{eqnarray}
J_{ADM}\left(\bar{\xi}_{i}\right)=\frac{1}{\kappa_{\text{Newton}}}\int_{S^{D-2}} & dS_{i} & \Big\{\bar{\xi}^{i}\partial_{j}h^{0j}-\bar{\xi}_{j}\partial^{i}h^{0j}\Big\},\label{eq:ADM_j}
\end{eqnarray}
 where $\bar{\xi}^{i}$ is the corresponding Killing vector. 

For asymptotically (anti)-de Sitter {[}(A)dS{]} spacetimes, the story
changes: the conserved charges are no longer simply geometric invariants
of the manifold, but \emph{theory-dependent} quantities. The parameters
of a theory enter the conserved charge expressions in such a way that
the charges are numerically scaled%
\footnote{In certain theories, linear combinations of the scaled ADM mass and
angular momentum are conserved charges, see topologically massive
gravity as an example \cite{DT-TMG}.%
} of the ADM charges as we shall see below. Taking the risk of being
pedantic, let us note that while the asymptotically flat Kerr black
hole solution has the same mass and the same angular momentum in all
geometric theories of gravity (in four dimensions), its asymptotically
(A)dS version Kerr-(A)dS black hole has different, numerically \emph{scaled}
masses and angular momenta for each theory to which it is a solution. 

Since at both low and high energies, general relativity is expected
to be modified for different reasons, one should build a procedure
to construct conserved charges in a given higher derivative theory.
What is perhaps also important is to find a formula that works in
all coordinates not just a specific one. 

The first generalization of the ADM mass was given by Abbott and Deser
\cite{Abbott} in cosmological Einstein's gravity for asymptotically
(A)dS spacetimes which reads in the notation of \cite{Deser_Tekin-PRL,Deser_Tekin-PRD}
as
\begin{eqnarray}
Q_{\text{Einstein}}^{\mu}(\bar{\xi})=\frac{1}{\kappa_{\text{Newton}}}\int_{\Sigma} & dS_{i} & \Big\{\bar{\xi}_{\nu}\bar{\nabla}^{\mu}h^{i\nu}-\bar{\xi}_{\nu}\bar{\nabla}^{i}h^{\mu\nu}+\bar{\xi}^{\mu}\bar{\nabla}^{i}h-\bar{\xi}^{i}\bar{\nabla}^{\mu}h\nonumber \\
 &  & +h^{\mu\nu}\bar{\nabla}^{i}\bar{\xi}_{\nu}-h^{i\nu}\bar{\nabla}^{\mu}\bar{\xi}_{\nu}+\bar{\xi}^{i}\bar{\nabla}_{\nu}h^{\mu\nu}-\bar{\xi}^{\mu}\bar{\nabla}_{\nu}h^{i\nu}+h\bar{\nabla}^{\mu}\bar{\xi}^{i}\Big\}.\label{ad}
\end{eqnarray}
For $\mu=0$, $Q^{0}\left(\bar{\xi}\right)$ gives the corresponding
energy or angular momentum once the background Killing vector $\bar{\xi}^{\mu}$
is specified. {[}Note that $Q^{i}\left(\bar{\xi}\right)$ is some
irrelevant current.{]} What is quite remarkable about (\ref{ad})
is that it not only works for asymptotically (A)dS spacetimes but
also for asymptotically flat ones. Thus, (\ref{ad}) combines the
ADM energy (\ref{eq:ADM_e}) and ADM angular momentum (\ref{eq:ADM_j})
in an \emph{arbitrary} coordinate system. (There is a small caveat
here: the coordinates should be sufficiently well-behaved at infinity,
see the Appendix of \cite{ST}.) The flat space limit of (\ref{ad})
in the cartesian coordinates is \cite{Tekin-Chern}

\begin{equation}
Q^{0}\left(\bar{\xi}\right)=\frac{1}{\kappa_{\text{Newton}}}\int_{S^{D-2}}dS_{i}\biggl(\bar{\xi}_{0}\Big\{\partial_{j}h^{ij}-\partial^{i}h_{jj}\Big\}+\bar{\xi}^{i}\partial_{j}h^{0j}-\bar{\xi}_{j}\partial^{i}h^{0j}\biggr).\label{eq:ADT_flat}
\end{equation}
as expected.

A second generalization of the ADM expression was carried out for
asymptotically (A)dS backgrounds in \cite{Deser_Tekin-PRL,Deser_Tekin-PRD}
for quadratic gravity theory with the action

\noindent 
\begin{equation}
I=\int d^{D}x\,\sqrt{-g}\left[\frac{1}{\kappa}\left(R-2\Lambda_{0}\right)+\alpha R^{2}+\beta R^{\mu\nu}R_{\mu\nu}+\gamma\left(R^{\mu\nu\rho\sigma}R_{\mu\nu\rho\sigma}-4R^{\mu\nu}R_{\mu\nu}+R^{2}\right)\right].\label{eq:Quadratic_action}
\end{equation}
We quote the result which shows that the effect of higher curvature
terms leads to a scaling of the charges computed in the cosmological
Einstein theory: 
\begin{align}
Q_{\text{quadratic}}^{\mu}(\bar{\xi})= & \left(\frac{1}{\kappa}+\frac{4\Lambda D}{D-2}\alpha+\frac{4\Lambda}{D-2}\beta+\frac{4\Lambda\left(D-3\right)\left(D-4\right)}{\left(D-1\right)\left(D-2\right)}\gamma\right)Q_{\text{Einstein}}^{\mu}(\bar{\xi}),\label{eq:Quad_charge}
\end{align}
where $Q_{\text{Einstein}}^{\mu}(\bar{\xi})$ is given in (\ref{ad})
(but with $\kappa_{\text{Newton}}=1$), and the effective cosmological
constant $\Lambda$ satisfies 
\begin{equation}
\frac{\Lambda-\Lambda_{0}}{2\kappa}+\left[\left(D\alpha+\beta\right)\frac{\left(D-4\right)}{\left(D-2\right)^{2}}+\gamma\frac{\left(D-3\right)\left(D-4\right)}{\left(D-1\right)\left(D-2\right)}\right]\Lambda^{2}=0.\label{quadratic}
\end{equation}

In this work, we will extend the discussion to generic $\mathcal{L}=\sqrt{-\left|g\right|}\, F\left(R_{\rho\sigma}^{\mu\nu}\right)$
theories. Save the theories which have $\left(\nabla_{\lambda_{1}}\dots\nabla_{\lambda_{i}}R_{\rho\sigma}^{\mu\nu}\right)^{n}$
type terms in the actions, our discussion below exhausts all the geometric
gravity theories.

The layout of the paper as follows: in the next section which is the
bulk of the paper, we construct the conserved charges of a generic
$F\left(R_{\rho\sigma}^{\mu\nu}\right)$ gravity by finding an equivalent
quadratic action that has the same $O\left(h\right)$ and $O\left(h^{2}\right)$
expansions as $F\left(R_{\rho\sigma}^{\mu\nu}\right)$ gravity. In
section III, we apply the formalism to the Born-Infeld gravity in
$2+1$ dimensions (BINMG). We use the mostly plus signature and the
Riemann and the Ricci tensors are defined as $\left[\nabla_{\mu},\nabla_{\nu}\right]V_{\lambda}=R_{\mu\nu\lambda}\,^{\sigma}V_{\sigma}$,\,\,
$R_{\mu\nu}\equiv R_{\mu\lambda\nu}\,^{\lambda}$.

\section{Charges of $F\left(R_{\rho\sigma}^{\mu\nu}\right)$ Gravity}

Our main task is to find the conserved charges of the following action
\begin{equation}
I=\int d^{D}x\,\sqrt{-\left|g\right|}\, F\left(R_{\rho\sigma}^{\mu\nu}\right),
\end{equation}
for asymptotically (A)dS spacetimes. The natural assumption on the
$F\left(R_{\rho\sigma}^{\mu\nu}\right)$ theory is that low energy
limit of the theory is Einstein's gravity. To this end, we can follow
the procedure given in \cite{Abbott,Deser_Tekin-PRL,Deser_Tekin-PRD}
which requires first to find the field equations and linearize them
about the (A)dS vacuum of the theory. Suppose the field equations
coupled to a matter source read as 
\begin{equation}
\Phi_{\mu\nu}(g,R,\nabla R,R^{2},...)=\kappa\tau_{\mu\nu},\label{generic}
\end{equation}
whose linearized forms symbolically become
\begin{equation}
{\cal O}(\bar{g})_{\mu\nu\alpha\beta}h^{\alpha\beta}=\kappa T_{\mu\nu},\label{ope}
\end{equation}
where $\bar{g}_{\mu\nu}$ satisfies $\Phi_{\mu\nu}(\bar{g},\bar{R},\bar{\nabla}\bar{R},\bar{R}^{2}...)=0$
and the deviation is defined as $h_{\mu\nu}\equiv g_{\mu\nu}-\bar{g}_{\mu\nu}$
and $T_{\mu\nu}$ includes all the higher order terms in $h_{\mu\nu}$
as well as the local matter source $\tau_{\mu\nu}$. The fact that
(\ref{ope}) is background covariantly conserved; i.e. $\bar{\nabla}_{\mu}T^{\mu\nu}=0$,
leads to the following globally conserved quantity 
\begin{equation}
Q\left(\bar{\xi}_{\nu}\right)\equiv\int_{\bar{\Sigma}}d^{D-1}y\,\sqrt{\bar{\gamma}}\bar{n}_{\mu}T^{\mu\nu}\bar{\xi}_{\nu}=\int_{\partial\bar{\Sigma}}d^{D-2}z\,\sqrt{\bar{\gamma}^{\left(\partial\bar{\Sigma}\right)}}\bar{n}_{\mu}\bar{\sigma}_{\nu}{\cal F}^{\mu\nu},\label{eq:Q}
\end{equation}
 where we have made use of the Stoke's theorem and assumed that a
background Killing vector $\bar{\xi}^{\mu}$ exists which leads to
$T^{\mu\nu}\bar{\xi}_{\nu}=\bar{\nabla}_{\nu}{\cal F}^{\mu\nu}$ where
${\cal F}^{\mu\nu}$ is an antisymmetric tensor. Here, $\bar{\gamma}$
is the induced metric on the hypersurface $\bar{\Sigma}$ which is
the spatial part of the spacetime manifold $\mathcal{M}$. $\partial\bar{\Sigma}$
is the boundary of $\bar{\Sigma}$. $\bar{n}^{\mu}$ is the normal
vector of the spatial $\left(D-1\right)$-dimensional hypersurface
$\bar{\Sigma}$, while $\bar{\sigma}^{\nu}$ is the normal vector
of the $\left(D-2\right)$-dimensional boundary $\partial\bar{\Sigma}$.
Note that written in (\ref{eq:Q}), $Q$ does not have any index,
it is the conserved charge. While this procedure is straightforward,
its actual execution for generic gravity, that is finding ${\cal F}^{\mu\nu}$
is rather tricky. Here, we follow another route, the so called equivalent
quadratic action formalism \cite{Hindawi,Gullu-UniBI,Gullu-AllUni3D,Sisman-AllUni},
and simplify the computation. This formalism boils down to finding
the equivalent quadratic action that has the same vacua and the same
linearized field equations as the $F\left(R_{\rho\sigma}^{\mu\nu}\right)$
theory under interest. From the above construction, it is clear that
conserved charges of the equivalent quadratic action and $F\left(R_{\rho\sigma}^{\mu\nu}\right)$
will be the same.

First, we recapitulate the construction of the equivalent quadratic
curvature action for a generic gravity theory defined with the Lagrangian
density $\mathcal{L}=\sqrt{-\left|g\right|}\, F\left(R_{\rho\sigma}^{\mu\nu}\right)$.
Note that the Riemann tensor is specifically chosen in the form with
two up and two down indices because any higher curvature term can
be constructed by solely the Riemann tensor in this form without any
need for the metric or its inverse. In addition, for the (A)dS background,
the Riemann tensor has the form $R_{\rho\sigma}^{\mu\nu}\sim\delta_{\rho}^{\mu}\delta_{\sigma}^{\nu}-\delta_{\sigma}^{\mu}\delta_{\rho}^{\nu}$
where the background metric $\bar{g}_{\mu\nu}$ and its inverse do
not appear, and this property simplifies the calculations of the equivalent
quadratic curvature action. 

To find the charges of the $F\left(R_{\rho\sigma}^{\mu\nu}\right)$
theory for asymptotically (A)dS spacetimes, the (A)dS vacua and the
linearized field equations for the $F\left(R_{\rho\sigma}^{\mu\nu}\right)$
theory should be determined through the $O\left(h\right)$ and $O\left(h^{2}\right)$
terms in the metric perturbation, $h_{\mu\nu}$, expansion of the
action $\int d^{D}x\,\mathcal{L}$. The up to $O\left(h^{2}\right)$
expansion of $F\left(R_{\rho\sigma}^{\mu\nu}\right)$, which is symbolically
\begin{equation}
F=F^{\left(0\right)}+\tau F^{\left(1\right)}+\tau^{2}F^{\left(2\right)}+O\left(\tau^{3}\right),\label{eq:Symbolic_h_exp_of_f}
\end{equation}
determines the $O\left(h\right)$ and $O\left(h^{2}\right)$ of $\int d^{D}x\,\mathcal{L}$
where we introduced a small parameter $\tau$. Therefore, any two
gravity theories defined with the functions, say, $F_{1}\left(R_{\rho\sigma}^{\mu\nu}\right)$
and $F_{2}\left(R_{\rho\sigma}^{\mu\nu}\right)$ have the same vacua
and the linearized field equations if and only if $F_{1}$ and $F_{2}$
have the same up to $O\left(h^{2}\right)$ expansions. Our aim is
to define a quadratic curvature gravity
\begin{equation}
f_{\text{quad-equal}}\left(R_{\rho\sigma}^{\mu\nu}\right)=\frac{1}{\kappa}\left(R-2\Lambda_{0}\right)+\alpha R^{2}+\beta R_{\nu}^{\mu}R_{\mu}^{\nu}+\gamma\left(R_{\rho\sigma}^{\mu\nu}R_{\mu\nu}^{\rho\sigma}-4R_{\nu}^{\mu}R_{\mu}^{\nu}+R^{2}\right),\label{eq:f_quad_equal-final_form}
\end{equation}
 with specific couplings to be determined below such that $F$ and
$f_{\text{quad-equal}}$ have the same up to $O\left(h^{2}\right)$
expansions. Hence, the gravity theories defined with the actions $\int d^{D}x\,\sqrt{-\left|g\right|}\, F\left(R_{\rho\sigma}^{\mu\nu}\right)$
and $\int d^{D}x\,\sqrt{-\left|g\right|}\, f_{\text{quad-equal}}\left(R_{\rho\sigma}^{\mu\nu}\right)$
are equivalent up to $O\left(h^{2}\right)$.

Having described the idea underlying the concept of the equivalent
quadratic curvature action, let us move on to the determination of
$f_{\text{quad-equal}}$ for a given $F$. Consider the Taylor series
expansion of $F$ in the curvature around the (A)dS background as
\begin{equation}
f\left(R_{\rho\sigma}^{\mu\nu}\right)=\sum_{i=0}^{\infty}\frac{1}{i!}\left[\frac{\partial^{i}F}{\partial\left(R_{\rho\sigma}^{\mu\nu}\right)^{i}}\right]_{\bar{R}_{\rho\sigma}^{\mu\nu}}\left(R_{\rho\sigma}^{\mu\nu}-\bar{R}_{\rho\sigma}^{\mu\nu}\right)^{i}.\label{eq:Taylor_exp_of_f}
\end{equation}
The simple but important point to notice is that the leading order
in the $h$ expansion of $\left(R_{\rho\sigma}^{\mu\nu}-\bar{R}_{\rho\sigma}^{\mu\nu}\right)$
is linear in $h$, that is 
\begin{equation}
R_{\rho\sigma}^{\mu\nu}-\bar{R}_{\rho\sigma}^{\mu\nu}=\tau\left(R_{\rho\sigma}^{\mu\nu}\right)_{\left(1\right)}+\tau^{2}\left(R_{\rho\sigma}^{\mu\nu}\right)_{\left(2\right)}+O\left(\tau^{3}\right);
\end{equation}
therefore, the leading order in the $h$ expansion of the $i^{\text{th}}$
order in (\ref{eq:Taylor_exp_of_f}) is $O\left(h^{i}\right)$ as
\begin{equation}
\left(R_{\rho\sigma}^{\mu\nu}-\bar{R}_{\rho\sigma}^{\mu\nu}\right)^{i}=\tau^{i}\left(R_{\rho\sigma}^{\mu\nu}\right)_{\left(i\right)}+O\left(\tau^{i+1}\right).
\end{equation}
With this observation, it is clear that the terms $F^{\left(0\right)}$,
$F^{\left(1\right)}$, and $F^{\left(2\right)}$ in the $h$ expansion
of $F$ involve contributions coming from only the orders $i\le2$
in (\ref{eq:Taylor_exp_of_f}). Therefore, the first three terms in
(\ref{eq:Taylor_exp_of_f}) determine the $O\left(h^{2}\right)$ expansion
of $F$. Then, $f_{\text{quad-equal}}$ can be defined as
\begin{equation}
f_{\text{quad-equal}}\left(R_{\rho\sigma}^{\mu\nu}\right)\equiv\sum_{i=0}^{2}\frac{1}{i!}\left[\frac{\partial^{i}F}{\partial\left(R_{\rho\sigma}^{\mu\nu}\right)^{i}}\right]_{\bar{R}_{\rho\sigma}^{\mu\nu}}\left(R_{\rho\sigma}^{\mu\nu}-\bar{R}_{\rho\sigma}^{\mu\nu}\right)^{i}.\label{eq:f_quad-equal_Riemann}
\end{equation}
As a side note, if the action for the higher curvature gravity solely
depends on the Ricci tensor as $\int d^{D}x\,\sqrt{-\left|g\right|}\, F\left(R_{\nu}^{\mu}\right)$,
one may use again (\ref{eq:f_quad-equal_Riemann}); however, the following
equivalent form of the $f_{\text{quad-equal}}\left(R_{\nu}^{\mu}\right)$
will be more convenient 
\begin{equation}
f_{\text{quad-equal}}\left(R_{\nu}^{\mu}\right)\equiv\sum_{i=0}^{2}\frac{1}{i!}\left[\frac{\partial^{i}F}{\partial\left(R_{\nu}^{\mu}\right)^{i}}\right]_{\bar{R}_{\nu}^{\mu}}\left(R_{\nu}^{\mu}-\bar{R}_{\nu}^{\mu}\right)^{i},\label{eq:f_quad-equal_Ricci}
\end{equation}
which follows from the same basic idea. Of course, for this case the
Gauss-Bonnet combination does not appear.

Once the equivalent quadratic curvature action, 
\begin{multline}
\int d^{D}x\,\sqrt{-g}\, f_{\text{quad-equal}}\left(R_{\rho\sigma}^{\mu\nu}\right)\\
=\int d^{D}x\,\sqrt{-g}\left[\frac{1}{\kappa}\left(R-2\Lambda_{0}\right)+\alpha R^{2}+\beta R_{\nu}^{\mu}R_{\mu}^{\nu}+\gamma\left(R_{\rho\sigma}^{\mu\nu}R_{\mu\nu}^{\rho\sigma}-4R_{\nu}^{\mu}R_{\mu}^{\nu}+R^{2}\right)\right],\label{eq:Equiv_quad_act}
\end{multline}
is found via (\ref{eq:f_quad-equal_Riemann}), then one can find the
(A)dS vacua and the charges for the asymptotically (A)dS spacetimes
by using the results of the generic quadratic curvature theory given
in \cite{Deser_Tekin-PRL,Deser_Tekin-PRD}. Therefore, there is no
need to either find the field equations or do an expansion in $h_{\mu\nu}$.

Suppose a specific theory is given, that is one knows the function
$F\left(R_{\rho\sigma}^{\mu\nu}\right)$, then let us summarize the
recipe to find the conserved charges of an asymptotically (A)dS solution
of this theory. One needs to calculate the following 
\begin{align}
\left[\frac{\partial F}{\partial R_{\rho\sigma}^{\mu\nu}}\right]_{\bar{R}_{\rho\sigma}^{\mu\nu}}R_{\rho\sigma}^{\mu\nu} & \equiv\zeta R,\label{eq:First_order}\\
\frac{1}{2}\left[\frac{\partial^{2}F}{\partial R_{\rho\sigma}^{\mu\nu}\partial R_{\lambda\gamma}^{\alpha\beta}}\right]_{\bar{R}_{\rho\sigma}^{\mu\nu}}R_{\rho\sigma}^{\mu\nu}R_{\lambda\gamma}^{\alpha\beta} & \equiv\alpha R^{2}+\beta R_{\sigma}^{\lambda}R_{\lambda}^{\sigma}+\gamma\left(R_{\rho\sigma}^{\mu\nu}R_{\mu\nu}^{\rho\sigma}-4R_{\nu}^{\mu}R_{\mu}^{\nu}+R^{2}\right),\label{eq:Second_order}
\end{align}
where $\zeta$, $\alpha$, $\beta$, $\gamma$ are to be determined
from these equations. $\alpha$, $\beta$ and $\gamma$ will appear
exactly in the equivalent quadratic action (\ref{eq:Equiv_quad_act}).
The other remaining two parameters of (\ref{eq:Equiv_quad_act}) follows
as
\begin{align}
\frac{1}{\kappa} & =\zeta-\left(\frac{4\Lambda}{D-2}\left(D\alpha+\beta\right)+\frac{4\Lambda\left(D-3\right)}{\left(D-1\right)}\gamma\right),\label{eq:kappa_eff}\\
\frac{\Lambda_{0}}{\kappa} & =-\frac{1}{2}F\left(\bar{R}_{\rho\sigma}^{\mu\nu}\right)+\frac{\Lambda D}{D-2}\zeta-\frac{2\Lambda^{2}D}{\left(D-2\right)^{2}}\left(D\alpha+\beta\right)-\frac{2\Lambda^{2}D\left(D-3\right)}{\left(D-1\right)\left(D-2\right)}\gamma.\label{eq:Lambda_0_eff}
\end{align}
Then, the gravitational charges of the $F\left(R_{\rho\sigma}^{\mu\nu}\right)$
theory is given as 
\begin{align}
Q_{F}^{\mu}(\bar{\xi})= & \left(\frac{1}{\kappa}+\frac{4\Lambda D}{D-2}\alpha+\frac{4\Lambda}{D-2}\beta+\frac{4\Lambda\left(D-3\right)\left(D-4\right)}{\left(D-1\right)\left(D-2\right)}\gamma\right)Q_{\text{Einstein}}^{\mu}(\bar{\xi}),\label{eq:Q_F}
\end{align}
where again $\alpha$, $\beta$, $\gamma$, $\kappa$ are to be found
from (\ref{eq:First_order}--\ref{eq:kappa_eff}), and the effective
cosmological constant $\Lambda$ satisfies (\ref{quadratic}).

\section{An Example: Charges of Born-Infeld Gravity (BINMG)}

As an application of the formalism developed in the previous section,
let us calculate the mass and angular momentum of the BTZ black hole
\cite{BTZ} for the BINMG theory \cite{Gullu-BINMG} defined with
the action
\begin{equation}
I_{\text{BINMG}}=-4m^{2}\int d^{3}x\,\left[\sqrt{-\det\left(g_{\mu\nu}+\frac{\sigma}{m^{2}}G_{\mu\nu}\right)}-\left(1-\frac{\lambda_{0}}{2}\right)\sqrt{-g}\right],\label{eq:det_BINMG_action}
\end{equation}
from which we can calculate the relevant quantities $F\left(\bar{R}_{\nu}^{\mu}\right)$,
(\ref{eq:First_order}), (\ref{eq:Second_order}) as

\begin{align}
F\left(\bar{R}_{\nu}^{\mu}\right) & =4m^{2}\left[\left(1-\frac{\lambda_{0}}{2}\right)-\left(1-\sigma\lambda\right)^{3/2}\right],\nonumber \\
\left[\frac{\partial F}{\partial R_{\beta}^{\alpha}}\right]_{\bar{R}_{\nu}^{\mu}}R_{\beta}^{\alpha} & =\sigma\left(1-\sigma\lambda\right)^{1/2}R,\label{eq:}\\
\frac{1}{2}\left[\frac{\partial^{2}F}{\partial R_{\sigma}^{\rho}\partial R_{\beta}^{\alpha}}\right]_{\bar{R}_{\nu}^{\mu}}R_{\sigma}^{\rho}R_{\beta}^{\alpha} & =\frac{1}{m^{2}}\left(1-\sigma\lambda\right)^{-1/2}\left(R_{\nu}^{\mu}R_{\mu}^{\nu}-\frac{3}{8}R^{2}\right),\nonumber 
\end{align}
where $\lambda\equiv\Lambda/m^{2}$ and $\sigma\lambda>1$ should
be satisfied. Therefore, one can simply read the effective parameters
of the equivalent quadratic action as
\begin{equation}
\zeta=\sigma\left(1-\sigma\lambda\right)^{1/2},\qquad\beta=-\frac{8}{3}\alpha=\frac{1}{m^{2}}\left(1-\sigma\lambda\right)^{-1/2}.
\end{equation}
Using this result in (\ref{eq:kappa_eff}) and (\ref{eq:Lambda_0_eff}),
one gets 
\begin{align}
\frac{1}{\kappa} & =\frac{\left(\sigma-\frac{\lambda}{2}\right)}{\sqrt{1-\sigma\lambda}},\nonumber \\
\frac{\Lambda_{0}}{\kappa} & =m^{2}\left[\lambda_{0}-2+\frac{1}{\sqrt{1-\sigma\lambda}}\left(2-\sigma\lambda-\frac{\lambda^{2}}{4}\right)\right].\label{eq:BINMG_kappa_Lambda_0}
\end{align}
 In \cite{Nam-Extended,Gullu-cfunc}, it was shown that the BTZ black
hole 
\begin{equation}
ds^{2}=-N^{2}dt^{2}+N^{-2}dr^{2}+r^{2}\left(N^{\phi}dt+d\phi\right)^{2},\label{eq:BTZ_metric}
\end{equation}
where
\begin{equation}
N^{2}\left(r\right)=-M+\frac{r^{2}}{\ell^{2}}+\frac{J^{2}}{4r^{2}},\qquad N^{\phi}\left(r\right)=-\frac{J}{2r^{2}},
\end{equation}
is a solution to BINMG theory under the condition 
\begin{equation}
\lambda=\sigma\lambda_{0}\left(1-\frac{\lambda_{0}}{4}\right),\quad\lambda_{0}<2.\label{eq:BTZ_BINMG}
\end{equation}
With out further due, by using (\ref{eq:Q_F}) the mass and the angular
momentum of the BTZ black hole in BINMG can be found as
\begin{equation}
E=\sigma\sqrt{1-\sigma\lambda}M,\qquad L=\sigma\sqrt{1-\sigma\lambda}J.\label{eq:E-L}
\end{equation}
Observe that as expected from (\ref{eq:Q_F}) the charges are scaled
yet their ratio is intact. This result matches with \cite{Nam-Extended}
where the charges were calculated using the black hole thermodynamics.

\section{Conclusions}

We have extended the Abbott-Deser-Tekin charge construction of linear
and quadratic gravity in asymptotically (A)dS spacetimes to generic
$F\left(R_{\rho\sigma}^{\mu\nu}\right)$ theory by finding a quadratic
action which has the same vacua and the linearized field equations
as the $F\left(R_{\rho\sigma}^{\mu\nu}\right)$ theory. We have applied
our method to the Born-Infeld gravity theory in $2+1$ dimensions
and confirmed the earlier calculations based on the thermodynamics
of the BTZ black hole.

\paragraph*{\textbf{Note added:}}

One day before this paper was submitted to the arXiv, \cite{Amsel}
appeared which deals with the same problem and reaches the same conclusions.

\section{Acknowledgments}

T.C.S. and B.T. are supported by the T{Ü}B\.{I}TAK Grant No. 110T339.

\end{document}